\documentclass[10pt,letterpaper]{article}
\usepackage{opex3}
\usepackage{graphicx}
\usepackage{epsfig}

\begin{document}

\title{Full-wave finite-difference time-domain simulation of electromagnetic cloaking structures}

\author{Yan Zhao*, Christos Argyropoulos, and Yang Hao}

\address{Queen Mary University of London, Mile End Road, London, E1 4NS, United Kingdom}

\email{yan.zhao@elec.qmul.ac.uk}

\begin{abstract}
This paper proposes a radial dependent dispersive finite-difference
time-domain method for the modelling of electromagnetic cloaking
structures. The permittivity and permeability of the cloak are
mapped to the Drude dispersion model and taken into account in
dispersive FDTD simulations. Numerical simulations demonstrate that
under ideal conditions, objects placed inside the cloak are
`invisible' to external electromagnetic fields. However for the
simplified cloak based on linear transformations, the back
scattering has a similar level to the case of a PEC cylinder without
any cloak, rendering the object still being `visible'. It is also
demonstrated numerically that the simplified cloak based on
high-order transformations can indeed improve the cloaking
performance.
\end{abstract}

\ocis{(230.3205) Invisibility cloaks; (000.4430) Numerical
approximation and analysis; (160.4760) Optical properties.}

\bibliographystyle{osajnl}

\section{Introduction}
The widespread interest in the invisibility of objects has led to
the recent development in electromagnetic cloaking structures.
Pendry \textit{et al.} proposed an electromagnetic material through
which electromagnetic fields can be controlled and manipulated to
propagate around its interior region like the flow of water
\cite{Pendry}, hence objects placed inside become `invisible' to
external electromagnetic fields. The proposed cloaking structure in
\cite{Pendry} requires inhomogeneous and anisotropic media, with
both permittivity and permeability independently controlled and
radially dependent. The magnitudes of the relative permittivity and
permeability of the perfect cloak are less than one, therefore it
cannot be constructed using naturally existing materials. The recent
development of metamaterials \cite{Veselago} (artificially
engineered structures with exotic electromagnetic properties which
cannot be obtained naturally) allows the control of material
parameters with freedom, and hence the construction of such cloaking
structures. However same as the negative-index metamaterials
\cite{Veselago}, the cloaking materials are inevitably dispersive
and therefore band-limited. Furthermore, the complete set of
material parameters proposed in \cite{Pendry} requires the control
of all the components of permittivity and permeability of the
material, making it difficult for practical realisation. It has been
proposed to use reduced sets of material parameters for both
transverse electric (TE) \cite{Cummer} and transverse magnetic (TM)
\cite{Cai} cases. The reduced set of material parameters for the TM
case eliminates the dependence on the material's magnetic property,
which is especially important for the realisation of the cloak in
the optical frequency range due to the absence of optical magnetism
in nature. However, considerable reflections occur because of the
impedance mismatch at the outer boundary of such a simplified cloak.
Under the assumption of the geometric optics, a high-order
transformation has been proposed in \cite{Cai2} to improve the
performance and minimise the scattering introduced by the cloak.

The development of Pendry's cloak is based on the coordinate
transformation technique \cite{Pendry,Leonhardt}, which has also
evoked other research topics such as the design of magnifying
perfect and super lenses \cite{Tsang}, the transformation media that
rotate electromagnetic fields \cite{ChenRotate}, the design of
reflectionless complex media for shifting and splitting optical
beams \cite{Rahm}, and the design of conformal antennas \cite{Luo}
etc. The spatial transformation technique has also been applied to
anslyse eccentric elliptic cloaks \cite{Kwon} and for acoustic
cloaking \cite{CummerAcoustic,ChenAcoustic}. Other theoretical
studies of the cloaking structure include the sensitivity of an
ideal cloak to tiny perturbations \cite{Ruan}, the performance of
the cylindrical cloaks with simplified material parameters
\cite{Yan,Isic,Zhang}, the realisation of cloaking using a
concentric layered structure of homogeneous isotropic materials
\cite{Huang}, the improvement of the cloaking performance using
soft-and-hard surface lining \cite{Greenleaf}, and the broadband
cloaking using sensors and active sources near the surface of a
region \cite{Miller} etc. The experimental demonstration of a
simplified cloak consisting of split-ring resonators (SRRs) has been
reported in microwave frequencies \cite{Schurig}. For the optical
frequency range, the cloak can be constructed by embedding silver
wires in a dielectric medium \cite{Cai}, or using a gold-dielectric
concentric layered structure \cite{Smolyaninov}.

It is worth mentioning that there exist different approaches to
render objects invisible, for example, by canceling the dipolar
scattering using plasmonic coatings \cite{Alu,Silveirinha}, and
using a left-handed media (LHM) coating \cite{Milton,Cory}. However,
the plasmonic coating approach is limited to the sub-wavelength
scale of the object, and the coating depends on the geometry and
material parameters of the object to be cloaked; the cloaking
performance using LHM coating is affected by the objects placed
inside. In comparison to these different approaches, Pendry's cloak
is more general and can be applied to objects with any dimensions
and under any wavelength condition.

The modelling of Pendry's invisible cloak has been performed by
using both analytical and numerical methods. Besides the widely used
coordinate transformation technique
\cite{Pendry,Leonhardt,Tsang,ChenRotate,Rahm,Luo,Kwon,CummerAcoustic,ChenAcoustic,Isic,SchurigRay},
a cylindrical wave expansion technique \cite{Ruan}, and a method
based on the full-wave Mie scattering model \cite{Chen} have also
been applied. In addition, the full-wave finite element method (FEM)
based commercial simulation package COMSOL Multiphysics$^{\rm TM}$
has been extensively used to model different cloaks and validate
theoretical predictions \cite{Cummer,Cai,Cai2,Kwon,Isic} due to its
ability of dealing with anisotropic and radial dependent material
parameters. While most of the numerical simulations have been
performed in the frequency domain, little attention has been paid to
the time-domain analysis of the cloaks. Frequency domain techniques
such as FEM may become inefficient when wideband solutions are
needed. So far the only time-domain analysis of the cloak is
presented in \cite{Weder} using a time-dependent scattering theory.
In this paper, we propose a dispersive finite-difference time-domain
(FDTD) method to deal with both the frequency and radial dependent
permittivity and permeability of the cloak. Due to its simplicity in
implementation and ability of treating anisotropic and inhomogeneous
materials, the FDTD method has been extremely popular for the
analysis of waveguides, transmission lines and antennas as well as
the calculation of dispersion diagrams of photonic crystals in both
microwave and optical communities. The aim of this paper is to
provide a full-wave FDTD solution to assist the analysis, design and
optimisation of the cloaking structures. To the authors' knowledge,
it is the first time that the FDTD method is implemented to model
cloaking structures.

\section{Dispersive FDTD modelling of the cloaking structure}
A complete set of material parameters of the ideal cloak is given by
\cite{Pendry}:
\begin{equation}
\varepsilon_r=\mu_r=\frac{r-R_1}{r},\qquad\varepsilon_\phi=\mu_\phi=\frac{r}{r-R_1},\qquad\varepsilon_z=\mu_z=\left(\frac{R_2}{R_2-R_1}\right)^2\frac{r-R_1}{r},
\label{eq_parameter_ideal}
\end{equation}
where $R_1$ and $R_2$ are the inner and outer radius of the cloak,
respectively. It can be easily identified from
(\ref{eq_parameter_ideal}) that the ranges of permittivity and the
permeability within the cloak are
$\varepsilon_r,\mu_r\in\left[0,\left(R_2-R_1\right)/R_2\right]$,
$\varepsilon_\phi,\mu_\phi\in\left[R_2/\left(R_2-R_1\right),\infty\right]$
and $\varepsilon_z,\mu_z\in\left[0,R_2/\left(R_2-R_1\right)\right]$.
Since the values of $\varepsilon_r$, $\mu_r$, $\varepsilon_z$ and
$\mu_z$ are less than one, same as the case for the LHMs, the cloak
cannot be directly modelled using the conventional FDTD method with
constant material parameters (at one particular location $r$).
However, one can map the material parameters using dispersive
material models e.g. Drude model (taking $\varepsilon_r$ as an
example)
\begin{equation}
\varepsilon_r(\omega)=1-\frac{\omega^2_p}{\omega^2-j\omega\gamma},
\label{eq_Drude}
\end{equation}
where $\omega_p$ and $\gamma$ are the plasma and collision
frequencies of the material, respectively. By varying the plasma
frequency, the radial dependent material parameters
(\ref{eq_parameter_ideal}) can be achieved. Note that in practise,
the plasma frequency of the material depends on the periodicity of
the SRRs \cite{Schurig} or wires \cite{Cai}, which varies along the
radial direction. Furthermore, different dispersion models (e.g.
Debye, Lorentz etc.) can be also considered for the modelling of the
cloak, which will lead to slightly different FDTD formula from the
following ones.

Since the conventional FDTD method \cite{Yee,Taflove} deals with
frequency-independent materials, the frequency-dependent FDTD method
is hence referred as the dispersive FDTD method
\cite{Luebbers,Gandhi,Sullivan}. For simplicity, in this paper, we
have implemented the dispersive FDTD method for a two-dimensional
(2-D) case and only three field components are non-zero: $E_x$,
$E_y$ and $H_z$. The modelled cloak is cylindrical and infinite
along $z$-direction ($\mu_r=\mu_\phi=\varepsilon_z=0$ in
(\ref{eq_parameter_ideal})), however an extension to a
three-dimensional (3-D) FDTD method to model a 3-D cloak
\cite{Pendry} is straightforward. There are also different
dispersive FDTD methods using different approaches to deal with
frequency-dependent material parameters: the recursive convolution
(RC) method \cite{Luebbers}, the auxiliary differential equation
(ADE) method \cite{Gandhi} and the $Z$-transform method
\cite{Sullivan}. Due to its simplicity, we have chosen the ADE
method for the modelling of the cloak.

The ADE dispersive FDTD method is based on Faraday's and Ampere's
Laws:
\begin{eqnarray}
\nabla\times\textbf{E}&=&-\frac{\partial\textbf{B}}{\partial t},
\label{eq_Maxwell_E}\\
\nabla\times\textbf{H}&=&\frac{\partial\textbf{D}}{\partial t},
\label{eq_Maxwell_H}
\end{eqnarray}
as well as the constitutive relations
$\textbf{D}=\varepsilon\textbf{E}$ and $\textbf{B}=\mu\textbf{H}$
where $\varepsilon$ and $\mu$ are expressed by
(\ref{eq_parameter_ideal}). Equations (\ref{eq_Maxwell_E}) and
(\ref{eq_Maxwell_H}) can be discretised following a normal procedure
\cite{Yee,Taflove} which leads to the conventional FDTD updating
equations:
\begin{eqnarray}
\textbf{B}^{n+1}&=&\textbf{B}^n-\Delta
t\cdot\widetilde{\nabla}\times\textbf{E}^{n+\frac{1}{2}},
\label{eq_Maxwell_B_approx}\\
\textbf{D}^{n+1}&=&\textbf{D}^n+\Delta
t\cdot\widetilde{\nabla}\times\textbf{H}^{n+\frac{1}{2}},
\label{eq_Maxwell_D_approx}
\end{eqnarray}
where $\widetilde{\nabla}$ is the discrete curl operator, $\Delta t$
is the discrete FDTD time step and $n$ is the number of the time
steps.

In addition, auxiliary differential equations have to be taken into
account and they can be discretised through the following steps. For
the conventional Cartesian FDTD mesh, since the material parameters
given in (\ref{eq_parameter_ideal}) are in cylindrical coordinates,
the coordinate transformation
\begin{equation}
\left[\begin{array}{cc}
\varepsilon_{xx} & \varepsilon_{xy}\\
\varepsilon_{yx} & \varepsilon_{yy}
\end{array}\right]=\left[\begin{array}{cc}
\varepsilon_r\cos^2\phi+\varepsilon_\phi\sin^2\phi & \left(\varepsilon_r-\varepsilon_\phi\right)\sin\phi\cos\phi\\
\left(\varepsilon_r-\varepsilon_\phi\right)\sin\phi\cos\phi &
\varepsilon_r\sin^2\phi+\varepsilon_\phi\cos^2\phi
\end{array}\right]
\label{eq_transform}
\end{equation}
is used. The tensor form of the constitutive relation is given by
\begin{equation}
\varepsilon_0\left[\begin{array}{cc}
\varepsilon_{xx} & \varepsilon_{xy}\\
\varepsilon_{yx} & \varepsilon_{yy}
\end{array}\right]\left[\begin{array}{c}
E_x\\E_y\end{array}\right]=\left[\begin{array}{c}
D_x\\D_y\end{array}\right]~~~\Leftrightarrow~~~\varepsilon_0\left[\begin{array}{c}
E_x\\E_y\end{array}\right]=\left[\begin{array}{cc}
\varepsilon_{xx} & \varepsilon_{xy}\\
\varepsilon_{yx} & \varepsilon_{yy}
\end{array}\right]^{-1}\left[\begin{array}{c}
D_x\\D_y\end{array}\right], \label{eq_constitutive}
\end{equation}
where
\begin{equation}
\left[\begin{array}{cc}
\varepsilon_{xx} & \varepsilon_{xy}\\
\varepsilon_{yx} & \varepsilon_{yy}
\end{array}\right]^{-1}=\frac{1}{\varepsilon_r\varepsilon_\phi}
\left[\begin{array}{cc}
\varepsilon_r\sin^2\phi+\varepsilon_\phi\cos^2\phi &
\left(\varepsilon_\phi-\varepsilon_r\right)\sin\phi\cos\phi\\
\left(\varepsilon_\phi-\varepsilon_r\right)\sin\phi\cos\phi &
\varepsilon_r\cos^2\phi+\varepsilon_\phi\sin^2\phi
\end{array}\right].
\label{eq_inverse}
\end{equation}
Note that the inverse of the permittivity tensor matrix exists only
when $\varepsilon_r\neq0$ and $\varepsilon_\phi\neq0$, which is not
the case for the inner boundary of the cloak. In our FDTD
simulations, we place a perfect electric conductor (PEC) cylinder
with radius equal to $R_1$ inside the cloak to guarantee the
validity of (\ref{eq_inverse}).

Substituting (\ref{eq_inverse}) into (\ref{eq_constitutive}) gives
\begin{equation}
\left\{\begin{array}{c}
\varepsilon_r\varepsilon_\phi\varepsilon_0E_x=\left(\varepsilon_r\sin^2\phi+\varepsilon_\phi\cos^2\phi\right)D_x+\left(\varepsilon_\phi-\varepsilon_r\right)\sin\phi\cos\phi D_y\\
\varepsilon_r\varepsilon_\phi\varepsilon_0E_y=\left(\varepsilon_r\cos^2\phi+\varepsilon_\phi\sin^2\phi\right)D_y+\left(\varepsilon_\phi-\varepsilon_r\right)\sin\phi\cos\phi
D_x
\end{array}\right.,
\label{eq_constitutive2}
\end{equation}
Express $\varepsilon_r$ in the Drude form of
(\ref{eq_parameter_ideal}), Eq. (\ref{eq_constitutive2}) can be
written as
\begin{equation}
\left\{\begin{array}{l}
\varepsilon_0\varepsilon_\phi\left(\omega^2-j\omega\gamma-\omega^2_p\right)E_x=\left[\left(\omega^2-j\omega\gamma-\omega^2_p\right)\sin^2\phi+\varepsilon_\phi\left(\omega^2-j\omega\gamma\right)\cos^2\phi\right]D_x\\
\qquad\qquad\qquad\qquad\qquad~~~~~+\left[\varepsilon_\phi\left(\omega^2-j\omega\gamma\right)-\left(\omega^2-j\omega\gamma-\omega^2_p\right)\right]\sin\phi\cos\phi D_y,\\
\varepsilon_0\varepsilon_\phi\left(\omega^2-j\omega\gamma-\omega^2_p\right)E_y=\left[\left(\omega^2-j\omega\gamma-\omega^2_p\right)\cos^2\phi+\varepsilon_\phi\left(\omega^2-j\omega\gamma\right)\sin^2\phi\right]D_y\\
\qquad\qquad\qquad\qquad\qquad~~~~~+\left[\varepsilon_\phi\left(\omega^2-j\omega\gamma\right)-\left(\omega^2-j\omega\gamma-\omega^2_p\right)\right]\sin\phi\cos\phi D_x.
\end{array}\right.
\label{eq_ExEy}
\end{equation}
Notice that $\varepsilon_\phi$ remains in (\ref{eq_ExEy}) because
its value is always greater than one (except at the inner surface of
the cloak) and can be directly used in conventional FDTD updating
equations \cite{Yee,Taflove}. Using an inverse Fourier transform and
the following rules:
\begin{equation}
j\omega\rightarrow\frac{\partial}{\partial
t},\qquad\omega^2\rightarrow-\frac{\partial^2}{\partial t^2},
\label{eq_inverse_Fourier}
\end{equation}
the first equation of (\ref{eq_ExEy}) can be rewritten in the time
domain as
\begin{eqnarray}
\lefteqn{\!\!\!\!\!\!\!\!\!\!\!\!\!\varepsilon_0\varepsilon_\phi\left(\frac{\partial^2}{\partial
t^2}+\gamma\frac{\partial}{\partial
t}+\omega^2_p\right)E_x=\left[\left(\frac{\partial^2}{\partial
t^2}+\gamma\frac{\partial}{\partial
t}+\omega^2_p\right)\sin^2\phi+\varepsilon_\phi\left(\frac{\partial^2}{\partial
t^2}+\gamma\frac{\partial}{\partial t}\right)\cos^2\phi\right]D_x}\nonumber\\
&&\qquad\qquad\qquad\qquad~~~+\left[\varepsilon_\phi\left(\frac{\partial^2}{\partial
t^2}+\gamma\frac{\partial}{\partial
t}\right)-\left(\frac{\partial^2}{\partial
t^2}+\gamma\frac{\partial}{\partial
t}+\omega^2_p\right)\right]\sin\phi\cos\phi D_y. \label{eq_Ex}
\end{eqnarray}

The FDTD simulation domain is represented by an equally spaced 3-D
grid with periods $\Delta x$, $\Delta y$ and $\Delta z$ along $x$-,
$y$- and $z$-directions, respectively. For discretisation of Eq.
(\ref{eq_Ex}), we use the central finite difference operators in
time ($\delta_t$ and $\delta^2_t$) and the central average operator
with respect to time ($\mu_t$ and $\mu^2_t$):
\begin{eqnarray}
\frac{\partial^2}{\partial t^2}\rightarrow\frac{\delta^2_t}{(\Delta
t)^2},\qquad\frac{\partial}{\partial
t}\rightarrow\frac{\delta_t}{\Delta
t}\mu_t,\qquad\omega^2_p\rightarrow\omega^2_p\mu^2_t,\nonumber
\label{eq_operator}
\end{eqnarray}
where the operators $\delta_t$, $\delta^2_t$, $\mu_t$ and $\mu^2_t$
are defined as in \cite{Hildebrand}:
\begin{eqnarray}
\delta_t\textbf{F}|^n_{m_x,m_y,m_z}&\equiv&\textbf{F}|^{n+\frac{1}{2}}_{m_x,m_y,m_z}-\textbf{F}|^{n-\frac{1}{2}}_{m_x,m_y,m_z},\\
\delta^2_t\textbf{F}|^n_{m_x,m_y,m_z}&\equiv&\textbf{F}|^{n+1}_{m_x,m_y,m_z}-2\textbf{F}|^n_{m_x,m_y,m_z}+\textbf{F}|^{n-1}_{m_x,m_y,m_z},\nonumber\\
\mu_t\textbf{F}|^n_{m_x,m_y,m_z}&\equiv&\frac{\textbf{F}|^{n+\frac{1}{2}}_{m_x,m_y,m_z}+\textbf{F}|^{n-\frac{1}{2}}_{m_x,m_y,m_z}}{2},\nonumber\\
\mu^2_t\textbf{F}|^n_{m_x,m_y,m_z}&\equiv&\frac{\textbf{F}|^{n+1}_{m_x,m_y,m_z}+2\textbf{F}|^n_{m_x,m_y,m_z}+\textbf{F}|^{n-1}_{m_x,m_y,m_z}}{4}.\nonumber
\label{eq_operators}
\end{eqnarray}
Here $\textbf{F}$ represents field components and $m_x,m_y,m_z$ are
indices corresponding to a certain discretisation point in the FDTD
domain. The discretised Eq. (\ref{eq_Ex}) reads
\begin{eqnarray}
\varepsilon_0\varepsilon_\phi\left[\frac{\delta^2_t}{(\Delta
t)^2}+\gamma\frac{\delta_t}{\Delta
t}\mu_t+\omega^2_p\mu^2_t\right]E_x=\left\{\left[\frac{\delta^2_t}{(\Delta
t)^2}+\gamma\frac{\delta_t}{\Delta
t}\mu_t+\omega^2_p\mu^2_t\right]\sin^2\phi\right.\nonumber\\
\left.+\varepsilon_\phi\left[\frac{\delta^2_t}{(\Delta
t)^2}+\gamma\frac{\delta_t}{\Delta
t}\mu_t\right]\cos^2\phi\right\}D_x+\left\{\varepsilon_\phi\left[\frac{\delta^2_t}{(\Delta
t)^2}+\gamma\frac{\delta_t}{\Delta t}\mu_t\right]\right.\nonumber\\
\left.-\left[\frac{\delta^2_t}{(\Delta
t)^2}+\gamma\frac{\delta_t}{\Delta
t}\mu_t+\omega^2_p\mu^2_t\right]\right\}\sin\phi\cos\phi D_y.
\label{eq_Ex1}
\end{eqnarray}
Note that in (\ref{eq_Ex1}), the discretisation of the term
$\omega^2_p$ of (\ref{eq_Ex}) is performed using the central average
operator $\mu^2_t$ in order to guarantee an improved stability; the
central average operator $\mu_t$ is used for the term containing
$\gamma$ to preserve the second-order feature of the equation.
Equation (\ref{eq_Ex1}) can be written as
\begin{eqnarray}
&&\varepsilon_0\varepsilon_\phi\left[\frac{E^{n+1}_x-2E^n_x+E^{n-1}_x}{(\Delta
t)^2}+\gamma\frac{E^{n+1}_x-E^{n-1}_x}{2\Delta
t}+\omega^2_p\frac{E^{n+1}_x+2E^n_x+E^{n-1}_x}{4}\right]\nonumber\\
&&=\sin^2\phi\left[\frac{D^{n+1}_x-2D^n_x+D^{n-1}_x}{(\Delta
t)^2}+\gamma\frac{D^{n+1}_x-D^{n-1}_x}{2\Delta
t}+\omega^2_p\frac{D^{n+1}_x+2D^n_x+D^{n-1}_x}{4}\right]\nonumber\\
&&\quad+\varepsilon_\phi\cos^2\phi\left[\frac{D^{n+1}_x-2D^n_x+D^{n-1}_x}{(\Delta
t)^2}+\gamma\frac{D^{n+1}_x-D^{n-1}_x}{2\Delta t}\right]\nonumber\\
&&\quad+\sin\phi\cos\phi\left\{\varepsilon_\phi\left[\frac{D^{n+1}_y-2D^n_y+D^{n-1}_y}{(\Delta
t)^2}+\gamma\frac{D^{n+1}_y-D^{n-1}_y}{2\Delta t}\right]\right.\nonumber\\
&&\quad\left.-\left[\frac{D^{n+1}_y-2D^n_y+D^{n-1}_y}{(\Delta
t)^2}+\gamma\frac{D^{n+1}_y-D^{n-1}_y}{2\Delta
t}+\omega^2_p\frac{D^{n+1}_y+2D^n_y+D^{n-1}_y}{4}\right]\right\}.
\label{eq_Ex2}
\end{eqnarray}
Therefore the updating equation for $E_x$ can be obtained as
\begin{equation}
E^{n+1}_x=\left[a_xD^{n+1}_x+b_xD^n_x+c_xD^{n-1}_x+d_x\overline{D_y}^{n+1}+e_x\overline{D_y}^n+f_x\overline{D_y}^{n-1}-\left(g_xE^n_x+h_xE^{n-1}_x\right)\right]/l_x.
\label{eq_Ex3}
\end{equation}
where the coefficients $a_x$ to $l_x$ are given by
{\footnotesize\begin{eqnarray}
a_x&\!\!\!\!\!=\!\!\!\!\!&\sin^2\phi\left[\frac{1}{(\Delta t)^2}+\frac{\gamma}{2\Delta t}+\frac{\omega^2_p}{4}\right]+\varepsilon_\phi\cos^2\phi\left[\frac{1}{(\Delta t)^2}+\frac{\gamma}{2\Delta t}\right],\nonumber\\
b_x&\!\!\!\!\!=\!\!\!\!\!&\sin^2\phi\left[-\frac{2}{(\Delta t)^2}+\frac{\omega^2_p}{2}\right]-\varepsilon_\phi\cos^2\phi\frac{2}{(\Delta t)^2},\nonumber\\
c_x&\!\!\!\!\!=\!\!\!\!\!&\sin^2\phi\left[\frac{1}{(\Delta t)^2}-\frac{\gamma}{2\Delta t}+\frac{\omega^2_p}{4}\right]+\varepsilon_\phi\cos^2\phi\left[\frac{1}{(\Delta t)^2}-\frac{\gamma}{2\Delta t}\right],\nonumber\\
d_x&\!\!\!\!\!=\!\!\!\!\!&\left\{\varepsilon_\phi\left[\frac{1}{(\Delta t)^2}+\frac{\gamma}{2\Delta t}\right]-\left[\frac{1}{(\Delta t)^2}+\frac{\gamma}{2\Delta t}+\frac{\omega^2_p}{4}\right]\right\}\sin\phi\cos\phi,\nonumber\\
e_x&\!\!\!\!\!=\!\!\!\!\!&\left\{\varepsilon_\phi\left[-\frac{2}{(\Delta t)^2}\right]-\left[-\frac{2}{(\Delta t)^2}+\frac{\omega^2_p}{2}\right]\right\}\sin\phi\cos\phi,\nonumber\\
f_x&\!\!\!\!\!=\!\!\!\!\!&\left\{\varepsilon_\phi\left[\frac{1}{(\Delta t)^2}-\frac{\gamma}{2\Delta t}\right]-\left[\frac{1}{(\Delta t)^2}-\frac{\gamma}{2\Delta t}+\frac{\omega^2_p}{4}\right]\right\}\sin\phi\cos\phi,\nonumber\\
g_x&\!\!\!\!\!=\!\!\!\!\!&\varepsilon_0\varepsilon_\phi\left[-\frac{2}{(\Delta
t)^2}+\frac{\omega^2_p}{2}\right],~~~h_x=\varepsilon_0\varepsilon_\phi\left[\frac{1}{(\Delta
t)^2}-\frac{\gamma}{2\Delta
t}+\frac{\omega^2_p}{4}\right],~~~l_x=\varepsilon_0\varepsilon_\phi\left[\frac{1}{(\Delta
t)^2}+\frac{\gamma}{2\Delta t}+\frac{\omega^2_p}{4}\right].\nonumber
\end{eqnarray}}
Following the same procedure, the updating equation for $E_y$ is
\begin{equation}
E^{n+1}_y=\left[a_yD^{n+1}_y+b_yD^n_y+c_yD^{n-1}_y+d_y\overline{D_x}^{n+1}+e_y\overline{D_x}^n+f_y\overline{D_x}^{n-1}-\left(g_yE^n_y+h_yE^{n-1}_y\right)\right]/l_y.
\label{eq_Ey}
\end{equation}
where the coefficients $a_y$ to $l_y$ are given by
{\footnotesize\begin{eqnarray}
a_y&\!\!\!\!\!=\!\!\!\!\!&\cos^2\phi\left[\frac{1}{(\Delta t)^2}+\frac{\gamma}{2\Delta t}+\frac{\omega^2_p}{4}\right]+\varepsilon_\phi\sin^2\phi\left[\frac{1}{(\Delta t)^2}+\frac{\gamma}{2\Delta t}\right],\nonumber\\
b_y&\!\!\!\!\!=\!\!\!\!\!&\cos^2\phi\left[-\frac{2}{(\Delta t)^2}+\frac{\omega^2_p}{2}\right]-\varepsilon_\phi\sin^2\phi\frac{2}{(\Delta t)^2},\nonumber\\
c_y&\!\!\!\!\!=\!\!\!\!\!&\cos^2\phi\left[\frac{1}{(\Delta t)^2}-\frac{\gamma}{2\Delta t}+\frac{\omega^2_p}{4}\right]+\varepsilon_\phi\sin^2\phi\left[\frac{1}{(\Delta t)^2}-\frac{\gamma}{2\Delta t}\right],\nonumber\\
d_y&\!\!\!\!\!=\!\!\!\!\!&\left\{\varepsilon_\phi\left[\frac{1}{(\Delta t)^2}+\frac{\gamma}{2\Delta t}\right]-\left[\frac{1}{(\Delta t)^2}+\frac{\gamma}{2\Delta t}+\frac{\omega^2_p}{4}\right]\right\}\sin\phi\cos\phi,\nonumber\\
e_y&\!\!\!\!\!=\!\!\!\!\!&\left\{\varepsilon_\phi\left[-\frac{2}{(\Delta t)^2}\right]-\left[-\frac{2}{(\Delta t)^2}+\frac{\omega^2_p}{2}\right]\right\}\sin\phi\cos\phi,\nonumber\\
f_y&\!\!\!\!\!=\!\!\!\!\!&\left\{\varepsilon_\phi\left[\frac{1}{(\Delta t)^2}-\frac{\gamma}{2\Delta t}\right]-\left[\frac{1}{(\Delta t)^2}-\frac{\gamma}{2\Delta t}+\frac{\omega^2_p}{4}\right]\right\}\sin\phi\cos\phi,\nonumber\\
g_y&\!\!\!\!\!=\!\!\!\!\!&\varepsilon_0\varepsilon_\phi\left[-\frac{2}{(\Delta
t)^2}+\frac{\omega^2_p}{2}\right],~~~h_y=\varepsilon_0\varepsilon_\phi\left[\frac{1}{(\Delta
t)^2}-\frac{\gamma}{2\Delta
t}+\frac{\omega^2_p}{4}\right],~~~l_y=\varepsilon_0\varepsilon_\phi\left[\frac{1}{(\Delta
t)^2}+\frac{\gamma}{2\Delta t}+\frac{\omega^2_p}{4}\right].\nonumber
\end{eqnarray}}
Note that the field quantities $\overline{D_x}$ in (\ref{eq_Ey}) and
$\overline{D_y}$ in (\ref{eq_Ex3}) are locally averaged values of
$D_x$ and $D_y$, respectively since the $x$- and $y$-components of
the field are in different locations in the FDTD domain. The
averaged values can be calculated using \cite{Lee}
\begin{eqnarray}
\overline{D_x}(i,j)&=&\frac{D_x(i,j)+D_x(i,j+1)+D_x(i-1,j)+D_x(i-1,j+1)}{4},\nonumber\\
\overline{D_y}(i,j)&=&\frac{D_y(i,j)+D_y(i+1,j)+D_y(i,j-1)+D_y(i+1,j-1)}{4},
\label{eq_Davg}
\end{eqnarray}
where $(i,j)$ is the coordinate of the field component.

The magnetic field $H_z$ can be calculated starting from the
constitutive relation $B_z=\mu_zH_z$, where $\mu_z$ is defined in
(\ref{eq_parameter_ideal}). Again we map the permeability using the
Drude model
\begin{equation}
\mu_z(\omega)=A\left(1-\frac{\omega^2_{pm}}{\omega^2-j\omega\gamma_m}\right),
\label{eq_Drudem}
\end{equation}
where we choose $A=2R_2/(R_2-R_1)$, and $\omega_{pm}$ and $\gamma_m$
are the magnetic plasma and collision frequencies of the material,
respectively. Following the above discretisation procedure, the
updating equation for $H_z$ can be derived as
\begin{eqnarray}
H^{n+1}_z&\!\!\!\!\!=\!\!\!\!\!&\frac{1}{A}\left\{\left[\frac{1}{\mu_0(\Delta
t)^2}+\frac{\gamma_m}{2\mu_0\Delta
t}\right]B^{n+1}_z-\frac{2}{\mu_0(\Delta
t)^2}B^n_z+\left[\frac{1}{\mu_0(\Delta
t)^2}-\frac{\gamma_m}{2\mu_0\Delta t}\right]B^{n-1}_z\right.\nonumber\\
&&\!\!\!\!\!\!\!\!\!\!\!\!\!\!\!\!\!\!\!\!\!\!\!\!\!\!\!\!\!\!\!\!\!\!\!\!\!\!\!
\left.+A\left[\frac{2}{(\Delta
t)^2}-\frac{\omega^2_{pm}}{2}\right]H^n_z-A\left[\frac{1}{(\Delta
t)^2}-\frac{\gamma_m}{2\Delta
t}+\frac{\omega^2_{pm}}{4}\right]H^{n-1}_z\right\}\Bigg/\left[\frac{1}{(\Delta
t)^2}+\frac{\gamma_m}{2\Delta t}+\frac{\omega^2_{pm}}{4}\right].
\label{eq_Hz}
\end{eqnarray}
Equations (\ref{eq_Maxwell_B_approx}), (\ref{eq_Maxwell_D_approx}),
(\ref{eq_Ex3}), (\ref{eq_Ey}) and (\ref{eq_Hz}) form an FDTD
updating equation set for the well-known leap-frog scheme
\cite{Yee}.

\begin{figure}[htbp]
\centering
\includegraphics[width=7cm]{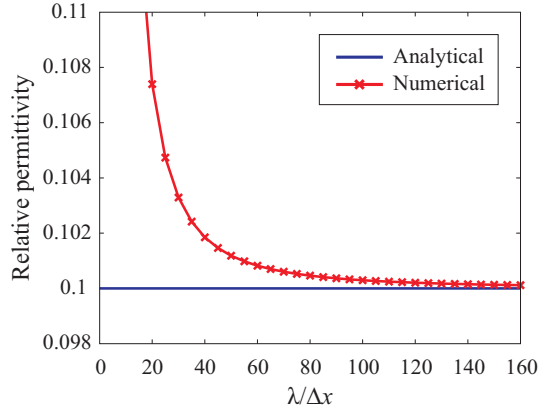}
\caption{The comparison between the analytical (\ref{eq_Drude}) and
numerical material parameters (\ref{eq_epsn}) for different FDTD
spatial resolutions for the case of $\varepsilon_r=0.1$ (lossless).}
\label{fig_parameters}
\end{figure}

Since the FDTD method is inherently a numerical technique, the
spatial as well as time discretisations have important effects on
the accuracy of simulation results. Also because the
permittivity/permeability is frequency dependent, one can expect a
slight difference between the analytical and numerical material
parameters due to the discrete time step in FDTD. In general, for
the modelling of conventional dielectrics with the relative
permittivity/permeability greater than one, a spatial resolution
(FDTD cell size) of $\Delta x<\lambda/10$ is required
\cite{Taflove}. However from our previous analysis \cite{ZhaoJOPA}
that for the modelling of metamaterials especially LHMs, the
numerical errors due to the time discretisation will cause spurious
resonances, hence a requirement of $\Delta x<\lambda/80$ is
necessary. For the case of the cloak, following the same approach as
in \cite{ZhaoJOPA} and for the case of plane-waves, substituting
\begin{equation}
\textbf{E}^{n}=\textbf{E}e^{jn\omega\Delta
t},\qquad\textbf{D}^{n}=\textbf{D} e^{jn\omega\Delta t},
\label{eq_central difference}
\end{equation}
into Eq. (\ref{eq_Ex2}), and comparing the resulting equation with
the first equation of (\ref{eq_constitutive2}) one can find that
$\widetilde{\varepsilon_\phi}$ (numerical permittivity along $\phi$
direction) has the exact analytical value, while
$\widetilde{\varepsilon_r}$ (numerical permittivity along radial
directions) takes the following form:
\begin{equation}
\widetilde{\varepsilon_r}=\varepsilon_0\left[1-\frac{\omega^2_p(\Delta
t)^2\cos^2\frac{\omega\Delta t}{2}}{2\sin\frac{\omega\Delta
t}{2}\left(2\sin\frac{\omega\Delta t}{2}-j\gamma\Delta
t\cos\frac{\omega\Delta t}{2}\right)}\right]. \label{eq_epsn}
\end{equation}
Note that Eq. (\ref{eq_epsn}) simplifies to the Drude dispersion
model (\ref{eq_Drude}) when $\Delta t\rightarrow0$. We plot the
comparison between the analytical (\ref{eq_Drude}) and numerical
relative permittivity (\ref{eq_epsn}) for the case of
$\varepsilon_r=0.1$ (lossless) in Fig.~\ref{fig_parameters}.
It is apparent that the conventional requirement of FDTD spatial
resolution does not guarantee the accuracy for the modelling of the
cloaks, and even for the case of $\Delta x=\lambda/40$, the
numerical error is still about 2\%.

With the expression of the numerical permittivity (\ref{eq_epsn})
available, one can correct the errors introduced by the discrepancy
between numerical and analytical material parameters. For example,
if the required permittivity is
$\varepsilon_r=\varepsilon'_r+j\varepsilon''_r$, after simple
derivations, the corrected plasma and collision frequencies can be
calculated as
\begin{equation}
\widetilde{\omega_p}^2=\frac{2\sin\frac{\omega\Delta
t}{2}\left[-2(\varepsilon'_r-1)\sin\frac{\omega\Delta
t}{2}-\varepsilon''_r\gamma\Delta t\cos\frac{\omega\Delta
t}{2}\right]}{(\Delta t)^2\cos^2\frac{\omega\Delta t}{2}},\quad
\widetilde{\gamma}=\frac{2\varepsilon''_r\sin\frac{\omega\Delta
t}{2}}{(\varepsilon'_r-1)\Delta t\cos\frac{\omega\Delta t}{2}}.
\label{eq_corrected}
\end{equation}

It is found from our FDTD simulations with $\Delta x=\lambda/35$
that without the correction of numerical material parameters, the
simulation becomes unstable before reaching the steady-state. The
cause of such an instability is currently under our investigation.
On the other hand, the correction of material parameters (both
permittivity and permeability) guarantees stable FDTD simulations.
Therefore in the following, the corrected material parameters
(\ref{eq_corrected}) are always used.

\section{Numerical results and discussion}
\begin{figure}[htbp]
\centering
\includegraphics[width=6.5cm]{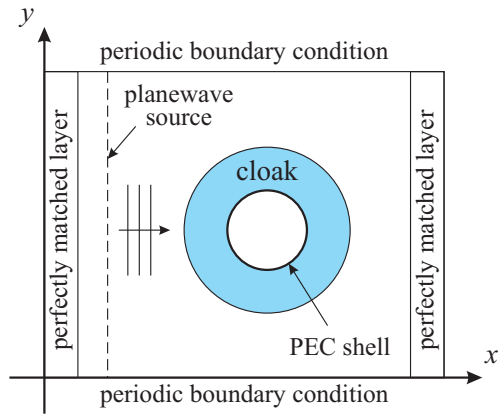}
\caption{A two-dimensional (2-D) FDTD simulation domain for the case
of plane-wave incidence on the cloak.} \label{fig_domain}
\end{figure}

The 2-D FDTD simulation domain is shown in Fig.~\ref{fig_domain}.
The following parameters are used in the simulations: FDTD cell size
$\Delta x=\Delta y=\lambda/150$ where $\lambda$ is the wavelength at
the operating frequency $f=2.0$ GHz, and the time step $\Delta
t=\Delta x/\sqrt{2}c$, chosen according to the Courant stability
criterion \cite{Taflove}. In the present paper, we assume the ideal
lossless case i.e. the collision frequency in (\ref{eq_Drude}) is
equal to zero ($\gamma=0$). The radial dependent plasma frequency
can be calculated using (\ref{eq_corrected}) with a given value of
$\varepsilon_r$ calculated from (\ref{eq_parameter_ideal}). The
radii of the cloak are: $R_1=0.1$ m and $R_2=0.2$ m. The computation
domain is truncated using Berenger's perfectly matched layer (PML)
\cite{Berenger} in $y$-direction to absorb waves leaving the
computation domain without introducing reflections, and terminated
with the periodic boundary conditions (PBCs) in $x$-direction for
the modelling of a plane-wave source. The plane-wave source is
implemented by specifying a complete column of FDTD cells using a
certain wave function (sinusoidal source in our case), see
Fig.~\ref{fig_domain}.

\begin{figure}[htbp]
\centering
\includegraphics[width=11cm]{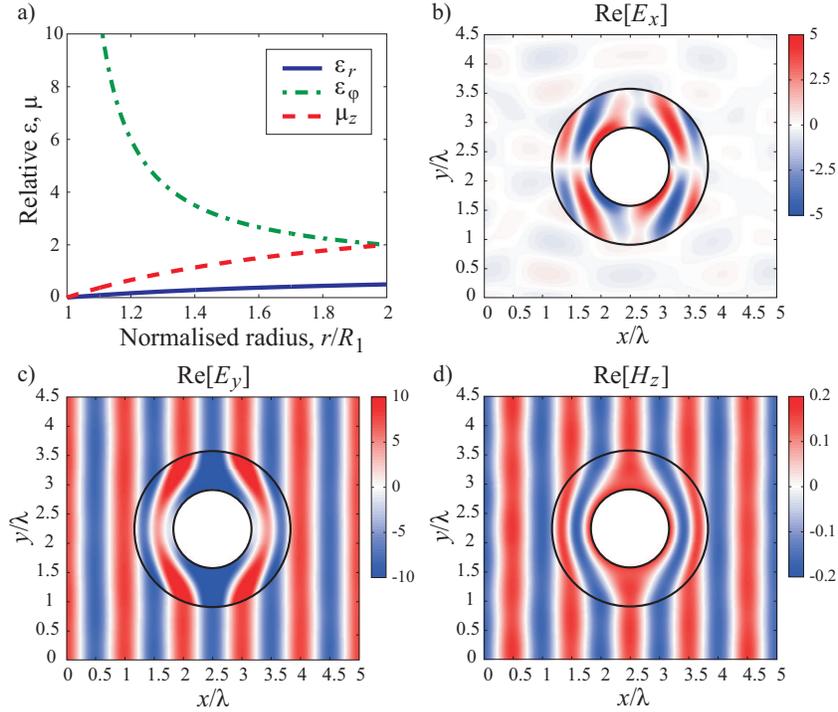}
\caption{(a) Material parameters for an infinite ideal cylindrical
cloak \cite{Cummer} where all $\varepsilon_r$, $\varepsilon_\phi$
and $\mu_z$ are radial dependent. (b), (c) and (d) Field
distributions from dispersive FDTD simulations of the cloak: (b)
$x$-component of the electric field, (c) $y$-component of the
electric field and (d) the magnetic field.} \label{fig_ideal}
\end{figure}

First we consider the ideal cloak, whose material parameters are
given by (\ref{eq_parameter_ideal}) with
$\mu_r=\mu_\phi=\varepsilon_z=0$. Figure~\ref{fig_ideal} shows the
distributions of the electric and magnetic field components
calculated from the dispersive FDTD simulation of the ideal cloak.
Note that only the central part of the simulation domain is shown
and the actual computation domain is larger. It is seen in
Fig.~\ref{fig_ideal}(b), (c) and (d) that the distribution is in
good agreement with those in \cite{Cummer,Cai} calculated using
frequency domain methods, although there are tiny ripples in the
magnetic field. In fact the field distribution presented in
\cite{Cummer} shows even stronger ripples which may be due to the
inadequate spatial resolution used in the calculation. These ripples
are purely numerical and will disappear when an extremely fine FDTD
grid is used. In our case, the slight ripples are also caused by the
staircase approximation of the circular surface of the cloak in the
Cartesian coordinate system. There are also non-zero scattered
fields in the $x$-component of the electric field outside the cloak
(see Fig.~\ref{fig_ideal}(b)), which is also a sign of numerical
errors because an ideal cloak does not introduce any scattering
outside the cloak. The staircase approximation only causes a very
small amount of numerical errors due to the fine mesh used in our
simulations. The staircase problem can be further alleviated to
improve the accuracy of simulations using a conformal scheme in
addition to the dispersive FDTD method i.e. using an effective
permittivity at material boundaries, as it was done for the case of
isotropic dispersive materials at planar \cite{ZhaoJOPA} and curved
boundaries \cite{Zhao}. However, the difficulty lies in the
anisotropy of the cloaking material, which leads to an eighth-order
differential equation to be discretised (the order of the
differential equation for the case of isotropic dispersive materials
is four \cite{Zhao}).

\begin{figure}[htbp]
\centering
\includegraphics[width=11cm]{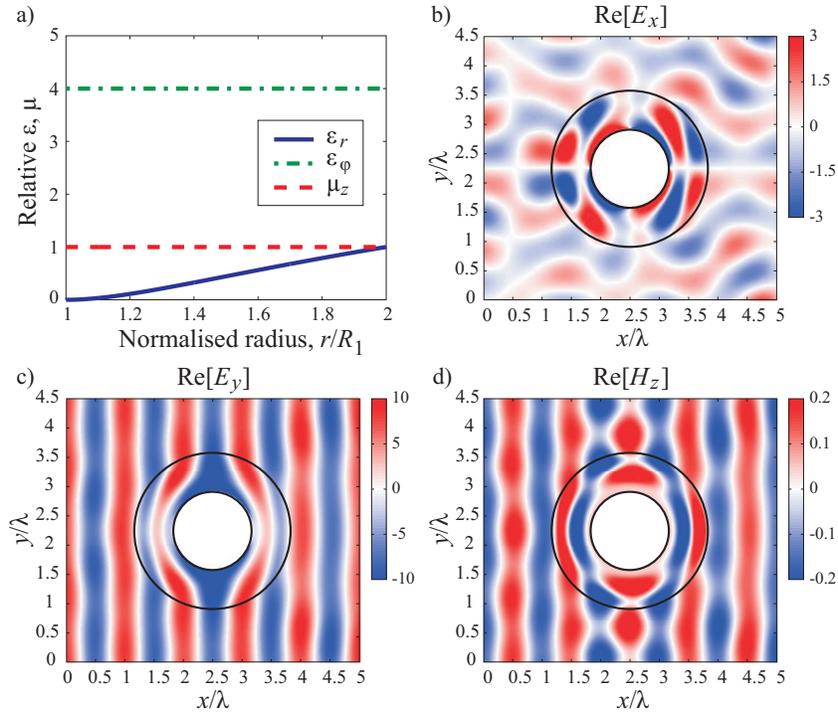}
\caption{(a) Material parameters for an infinite simplified
cylindrical cloak using a linear transformation \cite{Cai} where
only $\varepsilon_r$ is radial dependent. (b), (c) and (d) Field
distributions from dispersive FDTD simulations of the cloak: (b)
$x$-component of the electric field, (c) $y$-component of the
electric field and (d) the magnetic field.} \label{fig_reduced1}
\end{figure}

As mentioned previously, it is difficult to realise the ideal cloak
with all the components of permittivity and permeability being
radial dependent. Therefore it is proposed in \cite{Cai} that while
keeping the same wave trajectory, a reduced set of material
parameters also allows to construct a simplified cloak. The reduced
set of material parameters is given by \cite{Cai}:
\begin{equation}
\varepsilon_r=\left(\frac{R_2}{R_2-R_1}\right)^2\left(\frac{r-R_1}{r}\right)^2,\qquad
\varepsilon_\phi=\left(\frac{R_2}{R_2-R_1}\right)^2,\qquad\mu_z=1.
\label{eq_parameter_reduced}
\end{equation}
Following the same procedure, we have also modified the above
proposed dispersive FDTD method to model the reduced set of material
parameters and analysed its cloaking performance. The dimensions of
the simplified cloak remain the same as the ideal one. The field
distributions in the steady-state of the simulation are plotted in
Fig.~\ref{fig_reduced1}.
Such a cloak significantly reduces the complexity of practical
realisation since only $\varepsilon_r$ is radial dependent, as shown
in Fig.~\ref{fig_reduced1}(a). However, considerably reflections
occur due to the fact that the impedance matching at the outer
boundary of the simplified cloak is not satisfied anymore, see
Fig.~\ref{fig_reduced1}(b) and (d). Interestingly, the $y$-component
of the electric field is only affected slightly by the scattered
field. Note that here we only consider the simplified non-magnetic
cloak \cite{Cai}, the simplified cloak proposed in \cite{Cummer} can
be modelled in a similar way.

\begin{figure}[htbp]
\centering
\includegraphics[width=11cm]{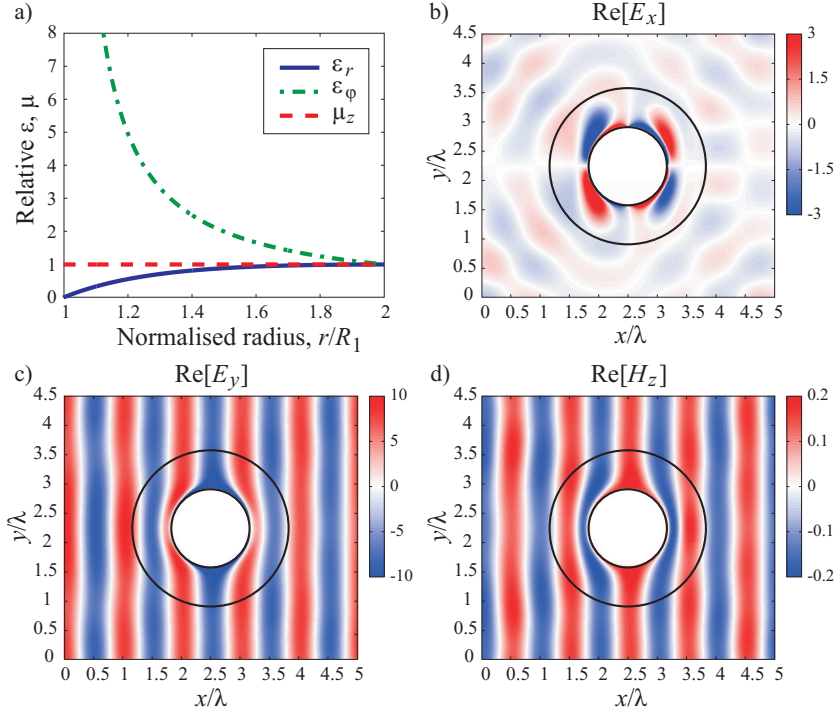}
\caption{(a) Material parameters for an infinite simplified
cylindrical cloak using a high-order transformation \cite{Cai2}
where only $\varepsilon_r$ and $\varepsilon_\phi$ are radial
dependent. (b), (c) and (d) Field distributions from dispersive FDTD
simulations of the cloak: (b) $x$-component of the electric field,
(c) $y$-component of the electric field and (d) the magnetic field.}
\label{fig_reduced2}
\end{figure}

The scattering from the simplified cloak
(\ref{eq_parameter_reduced}) due to the impedance mismatch can be
reduced by using an improved cloak based on a high-order
transformation \cite{Cai2}. The material parameters are given by
\cite{Cai2}
\begin{equation}
\varepsilon_r=\left(\frac{r'}{r}\right)^2,\qquad
\varepsilon_\phi=\left[\frac{\partial g(r')}{\partial
r'}\right]^{-2},\qquad\mu_z=1. \label{eq_parameter_reduced2}
\end{equation}
where $r=g(r')=\left[(R_1/R_2)(r'/R_2-2)+1\right]r'+R_1$. We have
also modelled such a cloak and plotted the field distributions in
Fig.~\ref{fig_reduced2}.
Its dimensions are kept the same as the previous two cases. In fact
the dimensions of this cloak is at its limit since it is required to
have $R_1/R_2<0.5$ to guarantee a monotonic transformation
\cite{Cai2}. The improved cloak adds an additional dependency of the
permittivity on the radius, as shown in Fig.~\ref{fig_reduced2}(a).
It is clear that indeed the high-order transformation based cloak
has an improved impedance on its outer boundary and hence
considerably reduces the scattered field. Notice that the wavefront
only starts to bend near the inner surface of the cloak, in
comparison to the case for the ideal cloak shown in
Fig.~\ref{fig_ideal}. This is due to the slow variance of the
impedance from the outer boundary towards inside the cloak.

\begin{figure}[htbp]
\centering
\includegraphics[width=11cm]{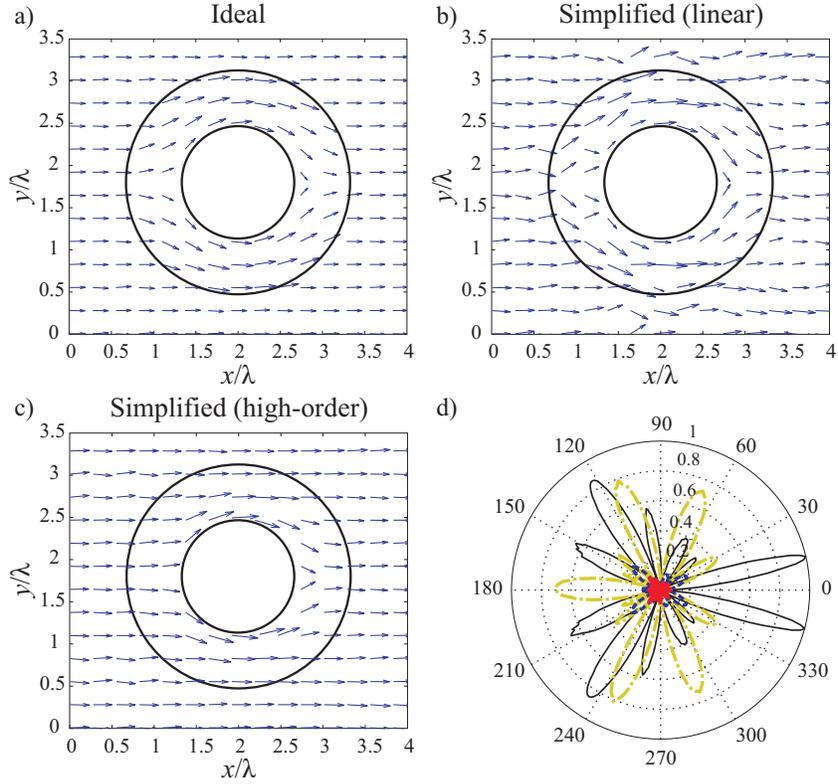}
\caption{Power flow diagrams for (a) the ideal cloak \cite{Cummer},
(b) the simplified cloak based on the linear transformation
\cite{Cai} and (c) the simplified cloak based on a high-order
transformation \cite{Cai2}. (d) Comparison of the scattering
patterns for different cloaks and for the case of the PEC cylinder
without cloak: black solid line - PEC cylinder; yellow dot-dashed
line - linear transformation based simplified cloak; blue dashed
line - high-order transformation based simplified cloak; red solid
line - ideal cloak.} \label{fig_poynting}
\end{figure}

For demonstration, we have also plotted the power flow diagrams as
well as the scattering patterns in Fig.~\ref{fig_poynting} for the
above ideal cloak, the simplified cloak based on a linear
transformation, and the simplified cloak based on a the high-order
transformation.
The power flow diagrams show that for the ideal case
(Fig.~\ref{fig_poynting}(a)), the wavefront enters the cloak
smoothly, bends around the central region and returns to its
original pattern after leaving the cloak. The cloak based on the
high-order transformation (Fig.~\ref{fig_poynting}(c)) shows a
similar pattern with a smooth bending of the wavefront near the
central region of the cloak. However due to reflections, the power
flow is disturbed before entering the cloak, while the wavefront
leaving the cloak has a relatively smoother distribution. For the
case of the cloak based on the linear transformation as shown in
Fig.~\ref{fig_poynting}(b), the waves do not strictly follow the
trend of the bending inside the cloak and propagate in arbitrary
directions. The consequence is that the external field is
significantly disturbed. This can be clearly identified from the
scattering patterns. The scattering pattern is calculated with the
reference to the free space case without any cloak or objects, and
then normalised to the scattering pattern of a PEC cylinder without
any cloak, see Fig.~\ref{fig_poynting}(d). For all the cloaks, the
scattering at the back of the cloak (relative to the direction of
wave incidence) is dramatically reduced. However the level of
scattered field for the linear cloak is almost the same as the case
for a PEC cylinder without any cloak, leading to the conclusion that
the object placed inside this simplified cloak can be detected from
its front side, similar to the conclusion drawn in \cite{Yan}. For
the high-order transformation based cloak, the scattering is reduced
by around four times comparing with the linear transformation based
one. Theoretically the ideal cloak has zero scattered field, however
the nonzero (small) values in Fig.~\ref{fig_poynting}(d) is mainly
caused by the staircase approximation in FDTD simulations, and will
approach zero when an extremely fine FDTD grid is used or a
conformal scheme is employed, as mentioned earlier.

\section{Conclusion}
In conclusion, we have proposed a dispersive FDTD method for the
modelling of the cloaking structure. The unusual material parameters
(the relative magnitudes of the permittivity and permeability are
less than one) are mapped to the Drude dispersion model, which is
then taken into account in FDTD simulations using an auxiliary
differential equation based method. The proposed method is
implemented in a two-dimensional case and three different
cylindrical cloaks are considered in our simulations: the ideal
cloak, the linear transformation based cloak and the high-order
transformation based cloak. It is found from the simulations that
the linear transformation based cloak introduces a level of back
scattering similar to the one of a PEC cylinder without the cloak,
causing the possibility of the object being detected. Such
scattering can be significantly reduced by using the high-order
transformation based cloak. In this paper we have only considered
lossless cloaks. The `ideal' cloak with a material loss of
$\tan\delta=0.1$ has been modelled in \cite{Cummer} using the finite
element method, and the case for $\tan\delta=0.01$, $0.1$ and $1$
modelled in \cite{Chen} using a full-wave Mie scattering theory.
Lossy cloaks can be also directly modelled using the above proposed
dispersive FDTD method by specifying a certain value for the
collision frequency in the Drude model for $\varepsilon_r$ and
defining a dielectric loss for $\varepsilon_\phi$.

\end{document}